\pdfoutput=1
\documentclass[a4paper,10pt,twocolumn]{article}

\usepackage[english]{babel}
\usepackage[utf8]{inputenc}
\usepackage[T1]{fontenc}

\usepackage{amsmath,amsfonts,amssymb}
\usepackage{graphicx}
\usepackage{xcolor}
\usepackage{tabularx}
\usepackage{booktabs}
\usepackage{subfig}
\usepackage{siunitx}
\usepackage{cite}
\usepackage[affil-it]{authblk}
\usepackage{abstract}

\usepackage{hyperref}
\hypersetup{colorlinks=true,urlcolor=oceanblue,citecolor=webgreen,linkcolor=blue,bookmarksopen=true,pdfauthor={M. Piazzi, J. Zemen, V. Basso},pdftitle={Ab-initio based analytical evaluation of entropy in magnetocaloric materials with first order phase transitions},pdfcreator={PDFLaTeX with hyperref package},pdfpagemode=UseOutlines,pdfpagelayout=SinglePage,pdfstartview=FitH}

\newcommand{\ud}{\mathop{}\!\mathrm{d}}

\graphicspath{{Figures/}}

\definecolor{oceanblue}{rgb}{0.125,0.125,0.9453125}
\definecolor{webgreen}{rgb}{0,0.3984375,0}

\begin{document}

\title{Kinetics of heat flux avalanches at the first order transition in La(Fe-Mn-Si)$_{13}$-H$_{1.65}$ compounds}

\author[1]{M Piazzi\thanks{Corresponding author.\\ \hspace*{0.55cm}E-mail address: \href{mailto:m.piazzi@inrim.it}{m.piazzi@inrim.it}}}
\author[1,2]{C. Bennati}
\author[1]{V. Basso}

\affil[1]{\small{Istituto Nazionale di Ricerca Metrologica, Strada delle Cacce 91, 10135 -- Torino, Italy}}
\affil[2]{\small {Dipartimento di Scienza Applicata e Tecnologia, Politecnico di Torino, C.so Duca degli Abruzzi 24, 10129 -- Torino, Italy}}

\date{}

\twocolumn[
\maketitle
\vspace{-1.1cm}
\begin{onecolabstract}
We study heat flux avalanches occurring at the first order transition in La(Fe-Mn-Si)$_{13}$-H$_{1.65}$ magnetocaloric material. As the transition is associated to the phase boundaries motion that gives rise to the latent heat, we develop a non equilibrium thermodynamic model. By comparing the model with experimental calorimetry data available for Mn=0.18, we find the values of the intrinsic kinetic parameter $R_L$, expressing the damping for the moving boundary interface, at different magnetic fields. We conclude that by increasing field, thus approaching the critical point, the avalanches increase in number and their kinetics is slowed down.    
\end{onecolabstract}]
\saythanks

\vspace*{0.2cm}

\section{Introduction}\label{intro}
The kinetics underlying the first order ferromagnetic (FM) to paramagnetic (PM) transitions of La(Fe-Si)$_{13}$ and related magnetocaloric materials is still an open issue \cite{Kuzmin-operatfreq, Lovell-dynamicsLaFeSi, Fujita-kineticsLaFeSi, Kuepferling-kinetics}. Novel aspects about this feature can be learnt by analyzing heat flux signals obtained through temperature scans at very low rates \cite{Bennati-kinetics, Bennati-unpub}. Indeed, the low scan rate allows to distinguish single heat flux avalanches associated to the microscopic individual processes occurring during the phase transition. Then, the characteristic times governing the behaviour of the avalanches are also those determining the kinetics of the transition. In particular, these indications may help to further improve the working frequency of magnetic refrigerators \cite{Kuzmin-operatfreq}. 

We develop a model, based on the non equilibrium thermodynamic theory of linear systems \cite{Callen-book}, in which the kinetics of first order transitions is related to the FM--PM phase boundaries motion. The model allows to describe single heat flux avalanches through an intrinsic kinetic parameter $R_L$ relating the change in latent heat to the difference between the transition temperature $T_t$ and the sample temperature $T_s$ (Sec.~\ref{theory}). We use experimental data provided by Peltier calorimetry under magnetic field performed at \SI[per-mode=symbol]{1}{\milli\kelvin\per\second} rate on LaFe$_{11.60}$Mn$_{0.18}$Si$_{1.22}$-H$_{1.65}$ compounds \cite{Basso-specheatLaFeSi}. In Sec.~\ref{results} we present the results of the comparison between the model and the experimental data, while in Sec.~\ref{discuss} we discuss them.

\section{Thermodynamic theory of heat flux avalanches}\label{theory}
We will describe the first order phase transition of the system as due to the motion of the boundaries between the FM and PM states. We assume these boundaries to be ideally thin interfaces characterized by a position $x(t)$, changing in time, and a surface area $A$. When the transition starts, the nucleation of the interface occurs at a temperature different from the equilibrium transition temperature $T_t$ because of the presence of local defects. Therefore, when transforming in heating from the FM to the PM phase, or viceversa in cooling from PM to FM, the nucleation event will be characterized by a critical temperature value $T_h$ higher or $T_c$ lower than $T_t$, respectively. The qualitative behaviour of the resulting path representing the sample temperature $T_s$ close to the transition region is sketched in \figurename~\ref{Fig1}.

\begin{figure*}[htbp]
\begin{minipage}[c]{.45\textwidth}
\centering
\includegraphics[width=\textwidth]{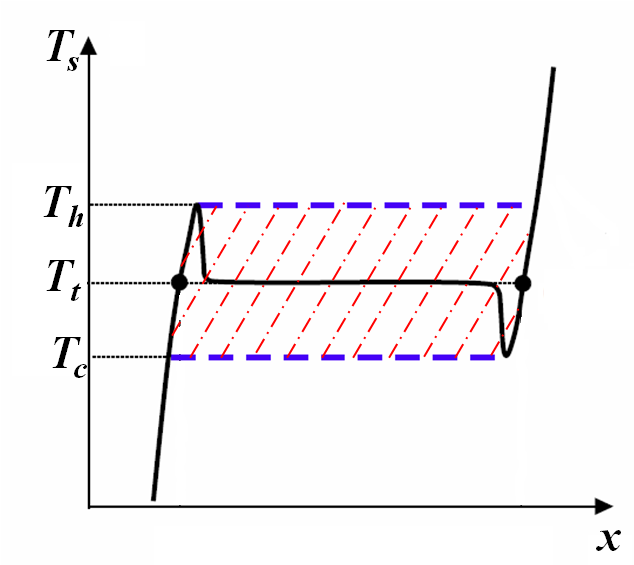}
\caption{Sketched behaviour of temperature $T_s$ in systems with first order transition and moving phase boundary position $x$. $T_h$, $T_c$, $T_t$ are the heating, cooling and equilibrium temperatures, respectively.}
\label{Fig1}
\end{minipage}\hspace{0.55cm}
\begin{minipage}[c]{.49\textwidth}
\renewcommand\figurename{Table}
\setcounter{figure}{0} 
\vspace*{2.2cm}
\footnotesize
\begin{tabularx}{\textwidth}{*{3}lp{2.25cm}l}
\toprule
$\boldsymbol{\mu_0 H}$ & $\boldsymbol{\tau_s}$ & $\boldsymbol{\Delta T_\text{\textbf{hys}}}$ & $\footnotesize{\boldsymbol{\Delta T_\text{\textbf{hys}}/\!\!\left(2\tau_sR_L\right)}}$ & $\boldsymbol{R_L}$ \\
{\bfseries (\si[detect-weight=true]{\tesla})} & {\bfseries (\si[detect-weight=true]{\second})} & {\bfseries (\si[detect-weight=true]{\kelvin})} & {\bfseries (\SI[per-mode=symbol,detect-weight=true]{E-3}[\times]{\watt\per\second})} & {\bfseries (\si[per-mode=symbol,detect-weight=true]{\kelvin\per\watt})}\\
\midrule
0 & 1.22 & 0.9 & 6.11 & 60\\
\midrule
0.5 & 2.95 & 0.6& 1.10 & 134\\
\midrule
1 & 2.78 & 0.4 & 0.56 & 129\\
\midrule
1.4 & 2.44 & 0.3 & 0.28 & 215\\
\bottomrule
\end{tabularx}
\caption{Values of the parameters $\tau_s$, $\Delta T_\text{hys}/\left(2\tau_s R_L\right)$ and $R_L$, appearing in Eq.~(\ref{linear_sol_qs}), at different magnetic fields $\mu_0 H$. Experimental values of $\Delta T_\text{hys}$ obtained from Ref.~\cite{Basso-specheatLaFeSi}.}
\label{Tab1}
\end{minipage}
\end{figure*}

With the above assumptions, the enthalpy change rate $\ud U_\text{L}/\ud t$, due to the motion of the full phase boundary interface, is given by
\begin{equation}
\frac{\ud U_\text{L}}{\ud t}=\Delta u_\text{L}A\frac{\ud x}{\ud t}
\label{latent_heat}
\end{equation} 
where $\Delta u_\text{L}$ is the enthalpy change per unit volume, representing the latent heat of the system.

To relate the velocity of the phase boundary interface $\ud x/\ud t$, appearing in Eq.~(\ref{latent_heat}), to the thermodynamic forces bringing the system back to equilibrium we employ the non equilibrium thermodynamic theory of linear systems \cite{Callen-book}. The fact that the transition may occur at temperature $T_s\neq T_t$ implies that the enthalpy variation $\ud U_\text{L}$ is written as $\ud U_\text{L}=T_s\left(\ud S-\sigma_s\ud t\right)$, where $\sigma_s$ is the entropy production rate, which is a definite positive term. Since the enthalpy change is also related to the entropy change $\ud S$ by $\ud U_\text{L}=T_t\ud S$, we obtain the following expression for $\sigma_s$:
\begin{equation}
\sigma_s=\left(\frac{1}{T_t}-\frac{1}{T_s}\right)\frac{\ud U_\text{L}}{\ud t}.
\label{entropy_prod_rate}
\end{equation} 
Eq.~(\ref{entropy_prod_rate}) states that the entropy production rate is the product of a distance-like term, i.e. $1/T_t-1/T_s$, and of a velocity-like term, i.e. $\ud U_\text{L}/\ud t$. The former represents how far from equilibrium the system is, while the latter describes the velocity of the process. Since we are considering that $T_s$ is close enough to $T_t$, we can safely assume that the displacement and the velocity terms are linked by a linear relation, obtaining this way the following equation:
\begin{align}
\notag \frac{\ud U_\text{L}}{\ud t} = & \alpha_s A T_t^2 \left(\frac{1}{T_t}-\frac{1}{T_s}\right)\\
&\simeq \alpha_s A\left(T_s-T_t\right)=\frac{1}{R_L}\left(T_s-T_t\right).
\label{relax_eq}
\end{align}
In Eq.~(\ref{relax_eq}), the last equality is justified by the assumption $T_s\simeq T_t$, the proportionality coefficient $\alpha_s$ is expressed in units \si{\watt\per\kelvin\per\metre\squared} and the intrinsic kinetic parameter $R_L={\left(\alpha_s A\right)}^{-1}$, which has the units of a thermal resistance, expresses the presence of damping processes for the phase boundary interface. By comparing Eq.~(\ref{relax_eq}) with Eq.~(\ref{latent_heat}) we can derive an expression for the velocity of the interface as: $\ud x/\ud t=\alpha_s T_t^2\left(1/T_t-1/T_s\right)/\Delta u_\text{L}\simeq\alpha_s\left(T_s-T_t\right)/\Delta u_\text{L}$.

The total enthalpy of the magnetocaloric material is the sum of the contributions from the latent heat, $U_\text{L}$, and from the specific heat, $U_\text{C}$: $U_e=U_\text{C}+U_\text{L}$. The contribution from the specific heat is taken as:
\begin{equation}
\frac{\ud U_\text{C}}{\ud t}=C_s\frac{\ud T_s}{\ud t} 
\label{rever_q}
\end{equation} 
where $C_s$ is the heat capacity of the region over which the transition takes place. 

By means of Eq.~(\ref{relax_eq}) and Eq.~(\ref{rever_q}), we have an expression of the heat flux $q_s$ exchanged with the surroundings: $q_s=\ud U_\text{C}/\ud t+\ud U_\text{L}/\ud t$. The time behaviour of $q_s$ is obtained by adding the effect of the measuring setup which is in contact with the material. A quasi-isothermal setup can be thought as a thermal bath at temperature $T_b$ in contact with the sample at temperature $T_s$ by a thermal resistance $R_s$. The heat $q_s(t)$ is then given by
\begin{equation}
q_s(t)=\frac{T_b(t)-T_s(t)}{R_s}.
\label{heat_flux_Peltier}
\end{equation} 

 By defining $t_0$ as the time at which the transition starts and $t_1$ as the time at which it ends, from Eqs.~(\ref{relax_eq}), (\ref{rever_q}), (\ref{heat_flux_Peltier}) we obtain the following couple of differential equations for $T_s(t)$:   
\begin{equation}
\left\{     
\begin{aligned}
C_s\frac{\ud T_s}{\ud t}&=\frac{T_b(t)-T_s(t)}{R_s}-\frac{T_s(t)-T_t}{R_L}\\
&\qquad\qquad\qquad\qquad\qquad\quad\, \text{ for } t_0\leq t\leq t_1\\[2pt]
C_s\frac{\ud T_s}{\ud t}&=\frac{T_b(t)-T_s(t)}{R_s} \qquad\qquad\quad\quad\:\text{ for } t> t_1.
\end{aligned}
\right.
\label{diff_eq_Ts}
\end{equation}
For an heating process, the $t_0\leq t\leq t_1$ regime describes the behaviour of the sample when it is absorbing the latent heat due to the phase boundary motion and the temperature $T_s$ decreases with respect to $T_b$. The $t> t_1$ regime describes instead what happens at the system when the transition is ended and $T_s$ relaxes towards $T_b$.

We have solved Eq.~(\ref{diff_eq_Ts}) in the case of a constant $T_b$, which is an assumption justified when dealing with temperature scans performed at very low rates. By assuming $T_b=T_s$ for $t\leq t_0$ and since at $t=t_0$, when the transition begins, the sample temperature $T_s$ reaches the threshold $T_h$, the constant $T_b$ value for an heating process is $T_b=T_b(t_0)=T_h$. Moreover, we have that also the initial conditions $T_0=T_s(t_0)=T_h$ and $T_1=T_s(t_1)$ hold. The solution $q_s(t)$ for an individual avalanche is:
\begin{equation}
q_s(t)=\left\{     
\begin{aligned}
&\frac{T_h-T_t}{R_s+R_L}\left(1-e^{-\tfrac{t-t_0}{\tau_{SL}}}\right) \quad\text{ for } t_0\leq t\leq t_1\\
&q_{s1}e^{-\tfrac{t-t_1}{\tau_s}} \qquad\qquad\quad\quad\quad\:\,\text{ for } t> t_1
\end{aligned}
\right.
\label{solution_qs}
\end{equation}
where $q_{s1}=q_s(t_1)$ is a constant factor. The time constants appearing in Eq.~(\ref{solution_qs}) are defined as $\tau_{SL}=\left(1/\tau_S+1/\tau_L\right)^{-1}$, $\tau_S=C_s R_s$ and $\tau_L=C_s R_L$. It is worth noting that \figurename~\ref{Fig2} shows that the growth part of an individual avalanche, being linear, is completed in a time $t_1-t_0$ much shorter than the time constant $\tau_{SL}$. Hence, we can approximate Eq.~(\ref{solution_qs}) by using the expansion $1-\exp{\left[-\left(t-t_0\right)/\tau_{SL}\right]}\simeq \left(t-t_0\right)/\tau_{SL}$, valid for $t-t_0\leq t_1-t_0\ll \tau_{SL}$. Furthermore, we observe that the time constant $\tau_{SL}$, being the parallel of the time constants $\tau_S$, $\tau_L$, is dominated by the smaller one between $\tau_S$ and $\tau_L$. 

\section{Comparison with experiments}\label{results}
\begin{figure}[htbp]
\centering
\includegraphics[width=.45\textwidth]{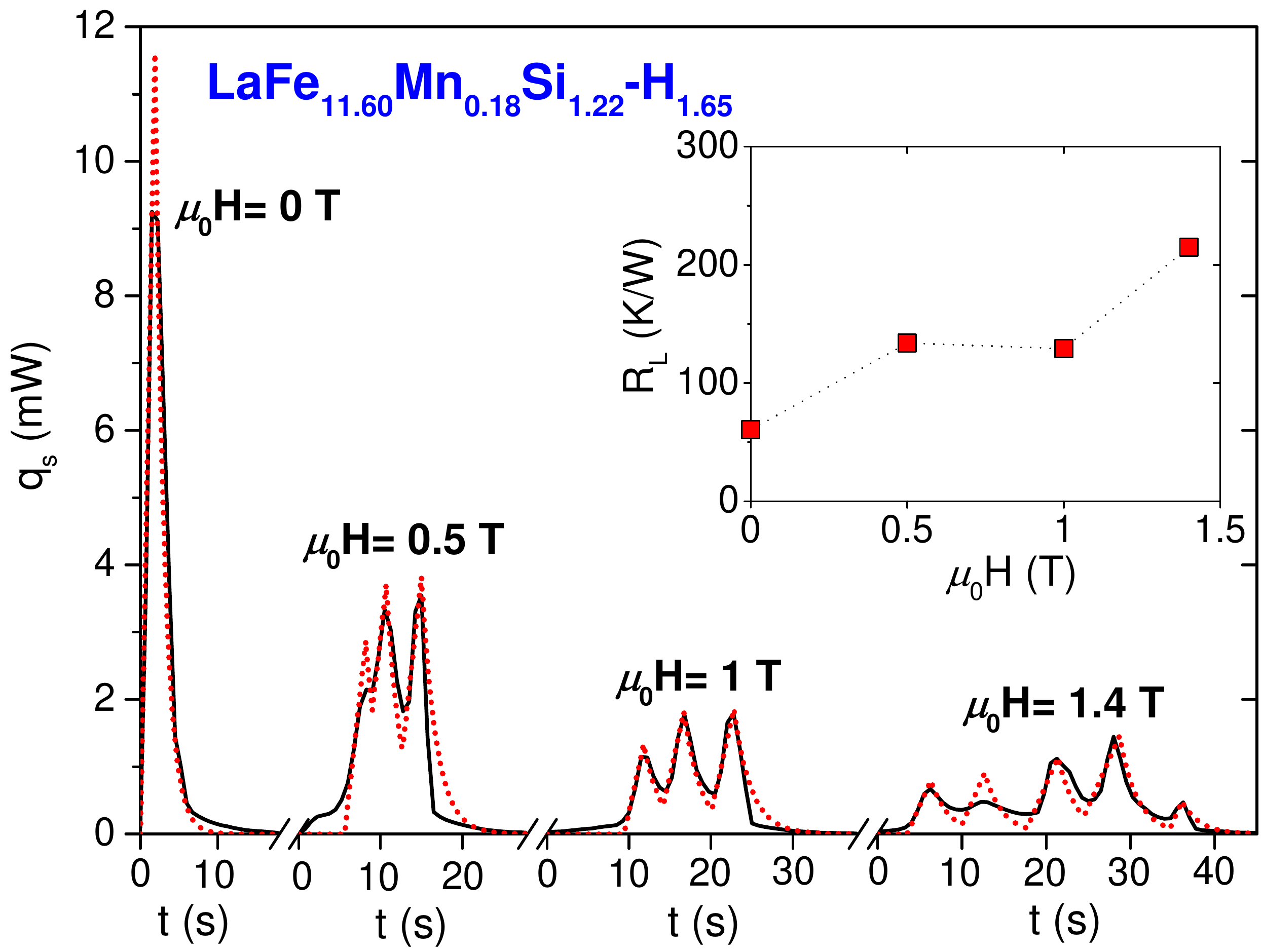}
\caption{Experimental (black solid lines) and modelled (dotted red lines) behaviour of heat flux $q_s$ vs. $t$ at different applied fields $H$, close to the first order transition. Measurements performed in heating, at \SI[per-mode=symbol]{1}{\milli\kelvin\per\second} rate, on a sample of mass $m=$ \SI{4.79}{\milli\gram} through Peltier calorimetry under magnetic field. Model curves obtained through Eq.~(\ref{linear_sol_qs}) with the parameters values reported in \tablename~\ref{Tab1}. Inset shows the intrinsic kinetic parameter $R_L$ of Eq.~(\ref{relax_eq}) at various $H$.}
\label{Fig2}
\end{figure}

\figurename~\ref{Fig2} shows the heat flux $q_s$ measured by Peltier calorimetry on a LaFe$_{11.60}$Mn$_{0.18}$Si$_{1.22}$-H$_{1.65}$ fragment of mass $m=$ \SI{4.79}{\milli\gram}, close to the transition, as a function of time $t$. The experimental data refer to heating scans performed at \SI[per-mode=symbol]{1}{\milli\kelvin\per\second} rate at increasing applied magnetic field $H$. It is clear the presence of subsequent heat flux avalanches increasing in number by increasing $H$. Each avalanche is characterized by an initial linear growth and an exponential decay whose slope and time constant change with the field. In particular, as soon as $H$ is getting closer to the critical field $\mu_0 H_c\sim$ \SI{2.2}{\tesla} ($\mu_0$ being the vacuum permeability) the transition becomes more and more of second order type \cite{Basso-specheatLaFeSi} and the avalanches start to be less clearly distinguishable from the background signal. The avalanches are related to the presence of latent heat, while the background is due to the reversible part of the specific heat of the system. 

 From the values of the resistivity of the silver varnish employed to contact the sample with the Peltier cell (\SI{2.5E3}{\watt\per\kelvin\per\metre\squared}) and of the contact area surface ($\sim$ \SI{1}{\milli\metre\squared}), we have $R_s\simeq$ \SI[per-mode=symbol]{400}{\kelvin\per\watt}. By comparing the model with the experimental data we have found $R_L< R_s$ (see \tablename~\ref{Tab1}), so we can safely assume $\tau_L < \tau_S$, so that $\tau_{SL}\simeq \tau_L$ and $R_s+R_L\simeq R_s$.     

By using the above observations, Eq.~(\ref{solution_qs}) takes the simpler form:
\begin{equation}
q_s(t)=\left\{     
\begin{aligned}
&\frac{T_h-T_t}{\tau_L R_s}\left(t-t_0\right)=\frac{\Delta T_\text{hys}}{2\tau_s R_L}\left(t-t_0\right)\\
&\qquad\qquad\qquad\qquad\quad\quad\;\,\text{ for } t_0\leq t\leq t_1\\[2pt]
&q_{s1}e^{-\tfrac{t-t_1}{\tau_s}} \qquad\qquad\qquad\quad\quad\,\,\text{ for } t> t_1
\end{aligned}
\right.
\label{linear_sol_qs}
\end{equation}
where we have used the approximation $T_h-T_t\approx \Delta T_\text{hys}/2$, with $\Delta T_\text{hys}=T_h-T_c$.

Eq.~(\ref{linear_sol_qs}) contains the free parameters $\tau_s$ and $\Delta T_\text{hys}/\left(2\tau_sR_L\right)$ that we have determined by comparison with the experimental data, trying to reproduce the qualitative behaviour of the avalanches. The result of the comparison is shown in \figurename~\ref{Fig2}. The key observation is that the initial growth and decay times, $t_0$ and $t_1$, have been chosen by concatenating subsequent individual avalanches characterized, for every fixed value of $H$, by the same slope $\Delta T_\text{hys}/\left(2\tau_sR_L\right)$ and the same time constant $\tau_s$. The values of these parameters, together with those of $\Delta T_\text{hys}(H)$ and $R_L$, are reported in \tablename~\ref{Tab1}. $R_L$ values for different $H$ are shown also in the inset of \figurename~\ref{Fig2}.

\section{Conclusions}\label{discuss}
We have applied the non equilibrium thermodynamic theory of linear systems to describe the heat flux avalanches observed by Peltier calorimetry temperature scans at low rates around the transition temperature of LaFe$_{11.60}$Mn$_{0.18}$Si$_{1.22}$-H$_{1.65}$ magnetocaloric material. We have found that a single avalanche shows an initial linear growth, describing the magnetic sample during the transition, and a subsequent exponential decay, corresponding to the system relaxation towards equilibrium. The decay behaviour is governed by a time constant $\tau_s\simeq$ \num{1}--\SI{2}{\second} increasing with the applied field $H$ and related only to the contact between the measuring setup and the sample. The growth depends instead on the intrinsic kinetic parameter $R_L$ introduced to describe the damping of the phase boundaries motion. Although we have analyzed only few experimental avalanches, the values of $R_L$ show indeed that the kinetic coefficient increases with $H$, indicating that the kinetics is slowed down by approaching the critical point of the transition. A detailed estimate of the value of the time constant $\tau_L$ associated to $R_L$ strictly depends on the typical length scales involving the sample region undergoing the transition and will be the subject of future work. Future investigations shall be also dedicated to understand how these outcomes may influence the operational frequency of magnetocaloric-based devices. 

\bibliographystyle{plain}
\bibliography{kinetics_biblio}

\end{document}